\documentclass[
    reprint,
    preprintnumbers,
    superscriptaddress,
    nofootinbib,
     amsmath,amssymb,
     aps,
     prd,
    floatfix, longbibliography
    ]{revtex4-2}

    \usepackage{graphicx}
    \usepackage[caption=false]{subfig}
    \usepackage[usenames,dvipsnames]{xcolor} 
    \usepackage{pgfplots}
    \usepackage[utf8]{inputenc}
    \usepackage{color}
    \usepackage{hyperref}
    \usepackage[normalem]{ulem} 
    \usepackage{physics}
    \usepackage{enumitem}
    \usepackage{comment}
    \usepackage{bm}
    \usepackage{booktabs}
\hypersetup{
  breaklinks = true,
  colorlinks   = true, 
  urlcolor     = Blue, 
  linkcolor    = Blue, 
  citecolor   = Blue 
}
    \allowdisplaybreaks[1]
    \graphicspath{{figures/}}

\usepackage{float}

\newcommand{\oxford}{Astrophysics, University of Oxford, DWB, Keble Road, Oxford OX1 3RH, United Kingdom}

\newcommand{\splitatcommas}[1]{%
  \begingroup
  \begingroup\lccode`~=`, \lowercase{\endgroup
    \edef~{\mathchar\the\mathcode`, \penalty0 \noexpand\hspace{0pt plus 1em}}%
  }\mathcode`,="8000 #1%
  \endgroup
}

\begin{document}

\title{Minimizing the tensor-to-scalar ratio in single-field inflation models}

\author{William J. Wolf}
\email{william.wolf@stx.ox.ac.uk}
\affiliation{\oxford}

\begin{abstract}
We revisit a class of simple single-field inflation models and demonstrate that they can readily produce a negligible tensor/scalar ratio $r$. Motivated by recent work suggesting the need to introduce higher order operators to stabilise unregulated potentials, as well as by work indicating that such terms can have significant effects on observable predictions, we explicitly construct corrected versions of the quadratic hilltop potential that are motivated by an effective field theory expansion. We employ Markov Chain Monte Carlo (MCMC) methods and optimization techniques to sample viable models and minimize $r$. We find that such potentials can readily lower $r$ values below projected CMB-S4 sensitivity, while still remaining within observable constraints on $n_s$. Furthermore, we find that the minimum $r$ reached for each order of the expansion considered is well-described by a power law $r_{min}(q) \propto q^{-B}$ before asymptoting to a value of $r_{min} \sim 10^{-11}$, where $q$ is the order to which the expansion of $V(\phi)$ is carried out.

\end{abstract}

\maketitle


\section{Introduction}\label{sec:introduction}

Inflation is currently the best theory we have to explain and describe the initial state of the early universe \cite{Guth:1980zm, Starobinsky:1980te, Albrecht:1982wi, Bardeen:1983qw, Linde:1981mu, Mukhanov:1981xt, Hawking:1982cz}. The theory of inflation proposes that the early universe experienced a period of accelerated, quasi-exponential expansion. During this period, inflationary dynamics produced a nearly flat and homogeneous universe, while quantum fluctuations in the inflaton field itself imprinted the universe with a characteristic spectrum of scalar density fluctuations/inhomogeneities that would seed the large-scale structure of the late-time universe.  The statistical features of this spectrum, observed in the Cosmic Microwave Background (CMB), largely align with generic inflationary predictions that suggest that these fluctuations should be adiabatic, Gaussian, and nearly scale-invariant \cite{Planck:2018jri, Planck:2018vyg, BICEPKeck:2022mhb}.

While the inflationary paradigm has been a spectacular success, seen a number of novel predictions confirmed, and justifiably become the dominant paradigm for modeling the early universe \cite{Guth:2013sya, Chowdhury:2019otk, Wolf:2023plp, Martin:2013tda, Martin:2024qnn, Wolf:2023jly}, 
the enthusiasm for inflation has been moderately tempered in recent years. This is no doubt partially due to the failure to detect observable B-modes, which have long hailed as a ``smoking-gun'' for inflation (see e.g.~\cite{Baumann:2009ds, CMBPolStudyTeam:2008rgp}; c.f.~\cite{Brandenberger:2011eq}) because they would inform us about another key prediction of inflation models: the tensor/scalar ratio $r$. This observable has been one of the primary parameters used to constrain inflationary models because it gives us direct information concerning the microphysics of inflation; i.e., its energy scale and the form of its potential. Furthermore, it has traditionally been held that standard, ``simple'' single-field inflationary models typically produce large, or at least observable, tensor/scalar ratios \cite{Easther:2021rdg}. Consequently, the null results have eliminated many of the simplest inflation models and inspired some to explore alternative approaches with renewed attention such as bouncing and string gas cosmologies (although inflation remains the dominant paradigm) \cite{Ijjas:2013vea, Ijjas:2015hcc, Steinhardt:2001st, Khoury:2001wf, Cai:2012va, Cai:2016thi, Easson:2011zy, Brandenberger:2016vhg, Brandenberger:2011et, Wolf:2022yvd, Martin:2024qnn, Chowdhury:2019otk, Guth:2013sya, Martin:2013nzq}. 

However, the answer to the question of what we can infer about the status of simple inflation models based on the detection or non-detection of $r$ is a subtle business. This of course depends upon one's definition of ``simple'', which, similarly to \cite{Stein:2022cpk}, we take to roughly be that the model consists of a single, canonical scalar field with a potential that can be approximated by an effective operator expansion (c.f.~\cite{Boyle:2005ug, Sousa:2023unz} for further discussions on possible definitions of ``simple'' in this context). To this point, the authors of \cite{Stein:2022cpk} provide a straightforward counter-example to the claim that simple single-field inflation models cannot produce a small tensor/scalar ratio: they construct a hilltop model with a potential described by a leading order quadratic term, and show that subleading order operators in the potential can induce an earlier end to inflation, and in doing so, lower $r$ ``arbitrarily''. 

This result intersects with some more general observations, especially emphasized by \cite{Kallosh:2019jnl} and further explored in \cite{Hoffmann:2021vty, Hoffmann:2022kod}, that:
\begin{enumerate}
    \item There are a number of inflation models in the literature with unregularized potentials that are unbounded from below.
    \item Producing a physically viable inflation model depends upon stabilizing, or ``correcting'', the potential. Otherwise, the universe would immediately re-collapse. 
    \item The observable predictions of inflation models can depend \textit{sensitively} on the nature of the correction terms that ensure inflation ends smoothly. 
\end{enumerate}
In light of these observations, it remains to be seen whether and to what extent an arbitrarily small tensor/scalar ratio can be realized within the construction of \cite{Stein:2022cpk}. 

In this paper, we answer this question by explicitly constructing models of inflation that can realize a vanishingly small $r$. To do so, we explore corrections to the quadratic hilltop model as suggested by \cite{Stein:2022cpk} to show how a small $r$ can be realized within this construction. Furthermore, while we find that these models can easily produce values for $r$ well below even the most optimistic observational sensitivities, we also find that $r$ can essentially be lowered arbitrarily, at least until its value to asymptote around $r \sim 10^{-11}$.

The paper proceeds as follows. Sec.~(\ref{sec:inflation}) describes the theory of inflation, slow-roll dynamics, and the observables associated with inflation models. Sec.~(\ref{sec:corrections}) discusses the necessity of correcting, or regularizing, unbounded inflationary potentials, and explores some of the ways that this has been done in the literature. This section also shows how the standard quadratic hilltop model can be corrected by considering the behavior of terms in the generic Taylor expansion of the potential. Sec.~(\ref{sec:predictions}) explores the observational predictions for $r$ and $n_s$ for this general model, and shows how $r$ can be lowered in a nearly arbitrary way before hitting a lower bound. Sec.~(\ref{sec:conclusion}) concludes.

\section{Inflationary slow-roll dynamics}\label{sec:inflation}
We begin with a discussion of the basic inflation scenario: a single, canonical scalar field $\phi$ minimally coupled to gravity in a homogeneous and isotropic FLRW background. 
The scale factor of the universe $a(t)$ evolves according to the Friedmann equations
\begin{align}
    H^2 &=\frac{1}{3 M_{\mathrm{Pl}}^2}\left(\frac{1}{2} \dot{\phi}^2+V(\phi)\right), \\
    \dot{H}  &=\frac{1}{2} \frac{\dot{\phi}^2}{M_{\mathrm{Pl}}^2},
\end{align}
where $H = \dot{a}/a$ and ${M_{\mathrm{Pl}}}$ is the Planck mass; and the scalar field $\phi$ evolves according to the Klein-Gordon equation in a Friedmann background
\begin{equation}
    \ddot{\phi}+3 H \dot{\phi}+V^{\prime}(\phi)=0,
\end{equation}
where $V^{\prime} = dV/d\phi$.

While these are the full dynamics, in order to produce the needed accelerated expansion, inflation occurs when the potential $V(\phi)$ is the primary driver of the dynamics. This permits us to work in the \textit{slow-roll} approximation. In this approximation, the kinetic term and the field acceleration are both vanishingly small $\dot{\phi}^2 = \ddot{\phi} \simeq 0$. This leads to the following system of simpler equations:
\begin{align}
    H^2 &\simeq \frac{V(\phi)}{3 M_{\mathrm{Pl}}^2}, \\
    3 H \dot{\phi} &\simeq -V^{\prime}(\phi).
\end{align}
The slow-roll approximation can be conveniently parameterized in terms of the \textit{slow-roll parameters} $\epsilon$ and $\eta$. 
\begin{align}
    \epsilon & \equiv \frac{M_{\mathrm{Pl}}^2}{2}\left(\frac{V^{\prime}}{V}\right)^2, \\
    \eta & \equiv M_{\mathrm{Pl}}^2 \frac{V^{\prime \prime}}{V}.
\end{align}
The accelerated expansion that inflation induces can only be sustained when these parameters are small, i.e., $\epsilon \ll 1$ and $\eta \ll 1 $, and inflation ends when $\epsilon (\phi_E) \simeq 1$.

We can also use this to calculate the number of \textit{e-folds} $N(\phi)$ between the end of inflation $E$ and some earlier initial time $I$ when the fluctuations exit the horizon and freeze.
\begin{equation}
    N(\phi) = \log (\frac{a_E}{a_I}) = \int_{\phi_{\mathrm{I}}}^{\phi_{\mathrm{E}}} \frac{d \phi}{M_{\mathrm{Pl}}} \frac{1}{\sqrt{2 \epsilon(\phi)}}.
\end{equation}
Inflation generally needs $N \simeq 60$ e-folds to successfully resolve the aforementioned fine-tuning puzzles as well as to be compatible with observations.

This theory provides an explanatorily satisfying solution to the fine-tuning puzzles that inspired its development and elegantly produces a flat, homogeneized universe. Where inflation really shines; however, is that the theory also explains and predicts how the universe generates the tiny inhomogeneities observed in the CMB that would later grow into the large-scale cosmic structure that we see today. Tiny quantum fluctuations in the inflaton field itself $\delta \phi$ produce density perturbations in the metric that inflationary dynamics then stretch and amplify over cosmological scales. Scalar density perturbations are produced with a power spectrum
\begin{align}
    \mathcal{P}_\phi = \left\langle|\delta \phi(k)|^2\right\rangle \propto A_s(k)^{n_s-1},
\end{align}
and amplitude
\begin{align}
    A_s =\frac{1}{8 \pi^2} \frac{1}{\epsilon} \frac{H^2}{M_{\mathrm{Pl}}},
\end{align}
where $k$ refers to the scale/wavenumber of the mode and $n_s$ is the so-called scalar spectrum index---a key observable that characterizes the scale-dependence of the fluctuation power spectrum which is defined as:
\begin{equation}
n_s(k)-1=\frac{d \ln \mathcal{P}_\phi}{d \ln k}.
\end{equation}

Additionally, inflation produces primordial gravitational waves, which are tensor perturbations that produce a distinctive B-mode polarization pattern. The amplitude of tensor perturbations is given by 
\begin{equation}
A_t=\frac{2}{\pi^2} \frac{H^2}{M_{\mathrm{Pl}}^2},
\end{equation}
which allows us to define the scalar/tensor ratio $r$ as
\begin{equation}
    r = \frac{A_s}{A_t}.
\end{equation}

The quantities $r$ and $n_s$ are the two primary observables used to constrain inflationary models. Furthermore, they can be directly calculated from the slow-roll parameters, leading to
\begin{align}
    n_s-1 &= -6 \epsilon\left(\phi_I\right)+2 \eta\left(\phi_I\right), \\
    r &= 16 \epsilon\left(\phi_I\right),
\end{align}
where quantities are evaluated when the modes exit the horizon at $\phi_I$.

\section{Unregularized inflation models and stabilising corrections}\label{sec:corrections}

\subsection{Observables are sensitive to corrections}

\begin{figure}
    \centering
    \begin{tikzpicture}
        \node at (0, 0) {\includegraphics[width=0.45\textwidth]{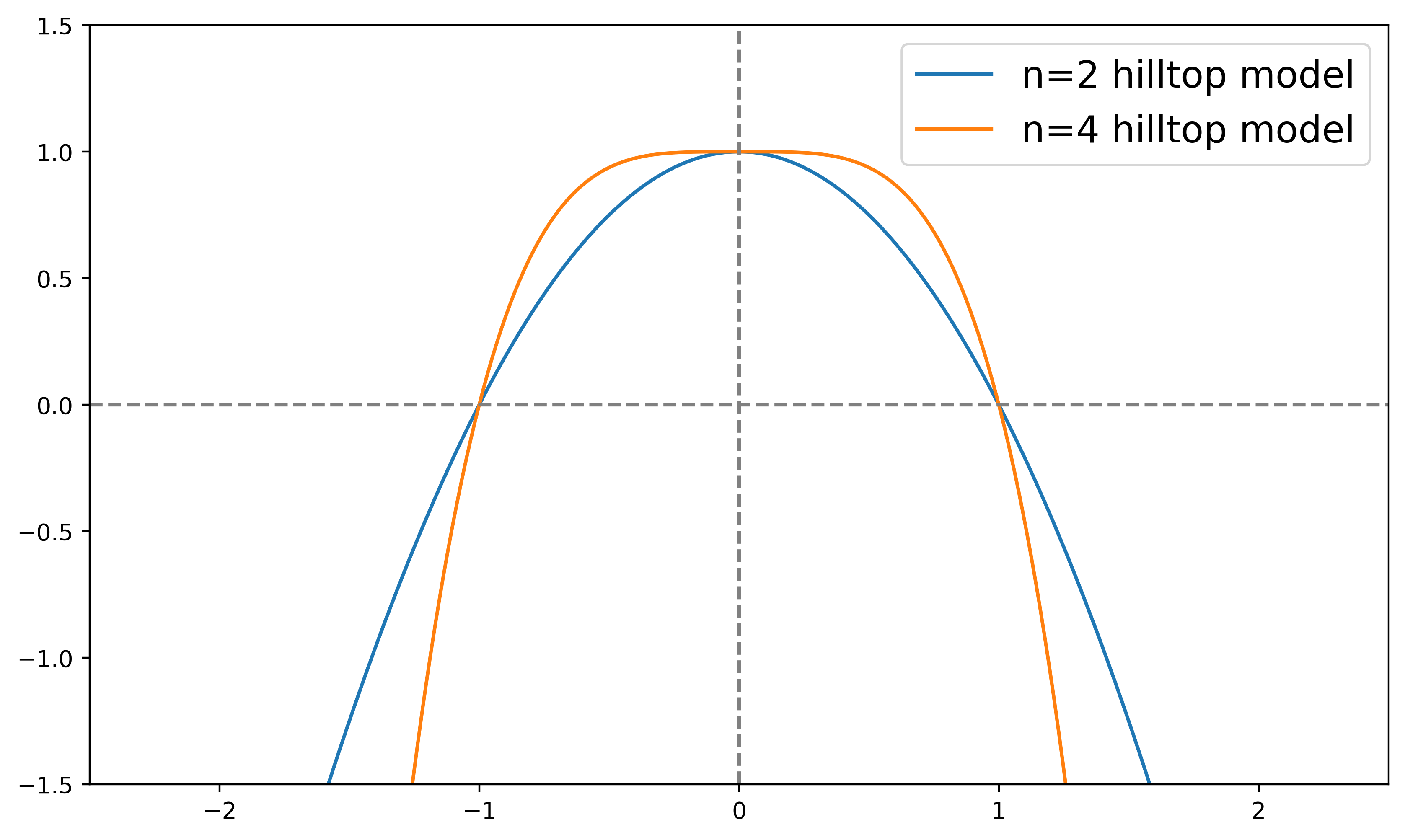}};
        \node[below] at (0.25, -2.50) {$\phi$};
        \node[rotate=90, above] at (-4.0, .15) {$V(\phi)$};
    \end{tikzpicture}
    \caption{Potentials $V(\phi)$ for quadratic ($n=2$) and quartic ($n=4$) hilltop models normalized to $V(0)=V_0=1$ with $\mu=1$. While the top of the hill is suitable for inflation, these potentials eventually become negative. These results are unphysical and would lead the post inflationary universe to immediately recollapse \cite{Felder:2002jk, Kallosh:2019jnl}. 
    }
    \label{hilltop}
\end{figure}

There are many models of inflation that have been considered which have potentials that are unbounded from below. Hilltop models are among these and will be the primary focus of this paper. This class of models has been investigated in a wide variety of contexts including inflation and dark energy (see e.g.~\cite{Boubekeur:2005zm, Kohri:2007gq, Dutta:2008qn, Wolf:2023uno, Lillepalu:2022knx, Dimopoulos:2020kol, Bostan:2018evz, Stein:2022cpk, Hoffmann:2021vty, Tzirakis:2007bf, Kinney:2006qm, German:2020rpn, Rashidi:2022ojt}), and is described by a potential of the following form,
\begin{equation}\label{hilltop potential}
    V(\phi) = V_0 \left[1 - \left(\frac{\phi}{\mu}\right)^n\right],
\end{equation}
where $V_0$ is the energy scale at the height of the potential, $\mu$ is a mass scale, and $n$ is an integer. While the region near the top of the potential clearly satisfies the basic criteria for a successful inflationary model, one can also easily see from Fig.~(\ref{hilltop}) that such potentials are unbounded from below. This is problematic because once $V < 0$, the universe will stop expanding, eventually begin to collapse, and fail to produce a viable cosmology \cite{Felder:2002jk, Kallosh:2019jnl}. Other potentials with this feature include D-brane inflation, radiatively corrected Higgs inflation, and exponential SUSY inflation (see e.g.~\cite{Martin:2013tda}). Due to these considerations, \cite{Kallosh:2019jnl} persuasively argues that \textit{any} model with this feature cannot be considered as a viable inflationary model; and furthermore, that for any of them to be valid they must be corrected such that they form stable minima that can secure a smooth exit from inflation. 

As it turns out, the nature of this stabilisation at the minima can dramatically affect observable predictions for $r$ and $n_s$ \cite{Martin:2013tda, Kallosh:2019jnl, Hoffmann:2022kod}. Thus, we cannot not naively assume the oft-cited $r-n_s$ predictions derived from from Eq.~(\ref{hilltop potential}) or others like it are representative of the predictions for viable versions of these models; rather, we need to actually construct stable models and compute the predictions.

\subsection{Stabilising the quadratic hilltop model}\label{sec:hilltop}

Inspired by the results of \cite{Stein:2022cpk}, we would like to explore constructing a stable version of the quadratic hilltop model ($n=2$), with the goal of seeing what freedom lies in higher order terms to lower $r$ ``arbitrarily''. 

How do we go about doing this? There are a few options. As \cite{Kallosh:2019jnl} notes, one can somewhat trivially stabilise the potential by introducing a finely-tuned correction term that very close approximates the form of Eq.~(\ref{hilltop potential}) for $\phi < \phi_{min}$, but then at $\phi_{min}$ becomes constant and turns sharply upwards for $\phi > \phi_{min}$. This preserves the predictions of Eq.~(\ref{hilltop potential}), but in an obviously ad-hoc and unsatisfying manner. However, we want to explore how to \textit{lower} $r$ while remaining viable in $n_s$, rather than merely retain the same predictions. 

An arguably more natural option would be to square Eq.~(\ref{hilltop potential}). This has been pursued for the $n=2$ quadratic hilltop squared inflation (also known as double-well inflation) \cite{Chowdhury:2019otk, Martin:2013tda} and $n=4$ quartic hilltop inflation \cite{Hoffmann:2021vty}. Yet, in both cases, squaring the potential has the effect of \textit{raising}, rather than lowering, $r$. For double well inflation, $r$ lies almost entirely outside data constraints, while quartic hilltop squared inflation remains viable, although its predictions for $r$ are pushed up notably.

Yet another option, which has been pursued in \cite{Hoffmann:2022kod} for the quartic hilltop model, is to add a polynomial term to stabilise the potential. For instance, one could stabilise the quadratic hilltop potential in the following way
\begin{equation}\label{polynomial}
V(\phi)=V_0\left[1-\left(\frac{\phi}{\mu}\right)^2+\alpha_p\left(\frac{\phi}{\mu}\right)^p\right], \quad p>2,
\end{equation}
where one can analytically work out the value of $\alpha_p$ to ensure that the potential reaches its minimum at zero. That is, 
\begin{equation}
    \phi_{vev} = \mu \left(\frac{2}{p \alpha_p}\right)^\frac{1}{p-2},
\end{equation}
and requiring that $V(\phi_{vev}) = 0$ allows us to solve for,
\begin{equation}
\alpha_p = \left( \frac{p}{2} -1 \right)^{\left(\frac{p}{2} -1\right)} \left(\frac{p}{2}\right)^{-\frac{p}{2}}.
\end{equation}

\begin{figure}
    \centering
    \begin{tikzpicture}
        \node at (0, 0) {\includegraphics[width=0.45\textwidth]{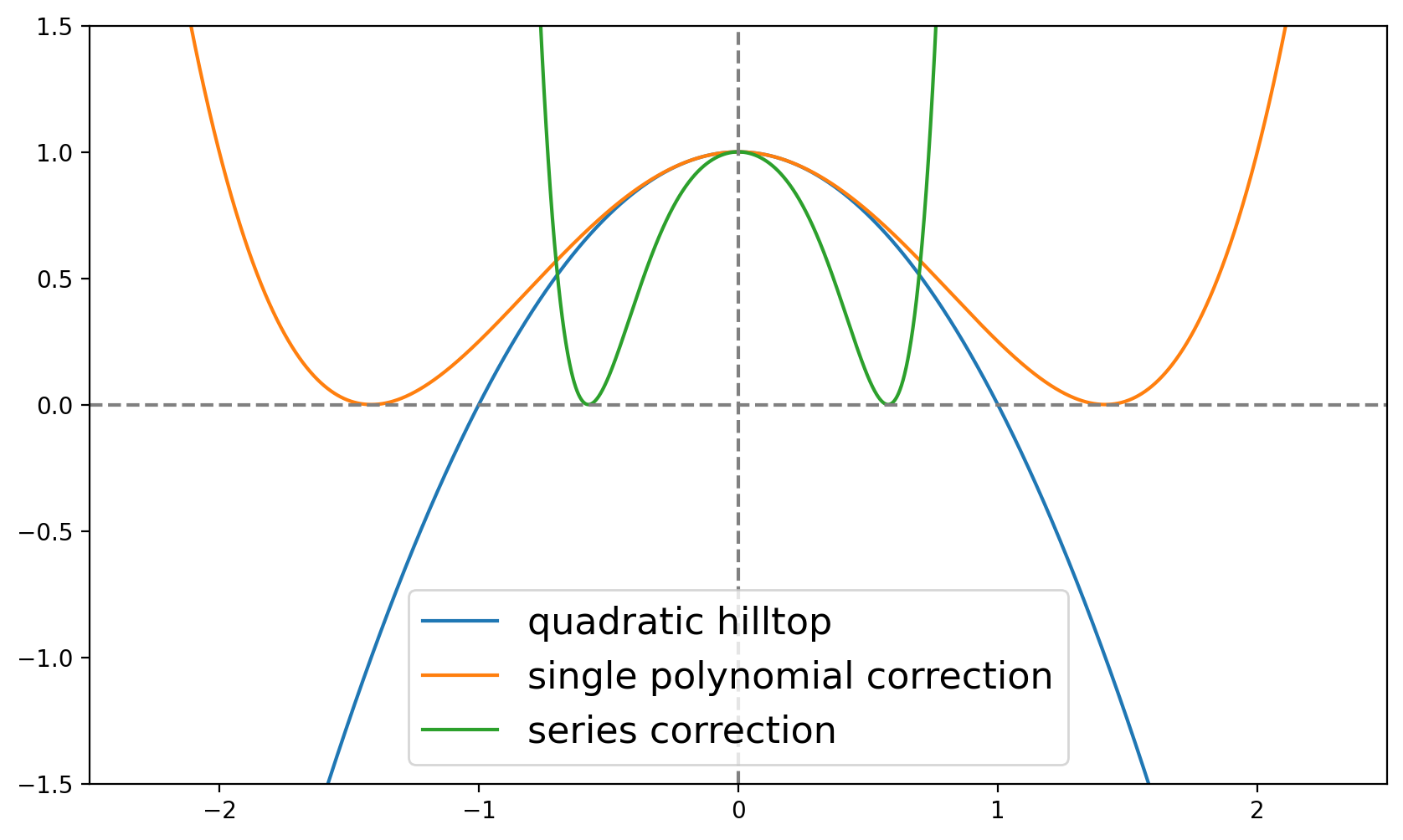}};
        \node[below] at (0.25, -2.5) {$\phi$};
        \node[rotate=90, above] at (-4.0, .15) {$V(\phi)$};
    \end{tikzpicture}
    \caption{The potentials $V(\phi)$ for the quadratic hilltop model (Eq.~(\ref{hilltop potential}) for $n=2$), a polynomial correction to the hilltop quadratic model (Eq.~(\ref{polynomial}) for $p=4$), and a more involved potential involving a series of multiple polynomial terms (given by Eq.~(\ref{series correction potential})). For visualization purposes, all potentials have been normalized to $V(0)=V_0=1$ and the mass scale chosen as $\mu = 1$. One can easily see from these potentials that the nature of the correction terms involved will have an impact on when inflation ends and the resulting observable predictions.}
    \label{corrected}
\end{figure}

However, just as \cite{Hoffmann:2022kod} found in the case of the quartic hilltop model, this has the effect of raising the predictions for $r$ and generally shifts curves in the $r-n_s$ plane up and towards the left, rather than down and towards the right. The higher order the single polynomial correction is, the closer that the inflationary observables will match the uncorrected model because the correction term will have less of an effect on the potential during the stages at which inflation is occurring; consequently though, adding a single polynomial stabilising term will not lower $r$. This is partially because simply adding adding a single correcting term will have the effect of making inflation end \textit{later}, whereas the results of \cite{Stein:2022cpk} indicate that the freedom to lower $r$ is at least somewhat tied to the ability to reduce $\phi_E$. See Fig.~(\ref{corrected}) where this potential is depicted for $p=4$.

\begin{figure}
    \centering
    \begin{tikzpicture}
        \node at (0, 0) {\includegraphics[width=0.45\textwidth]{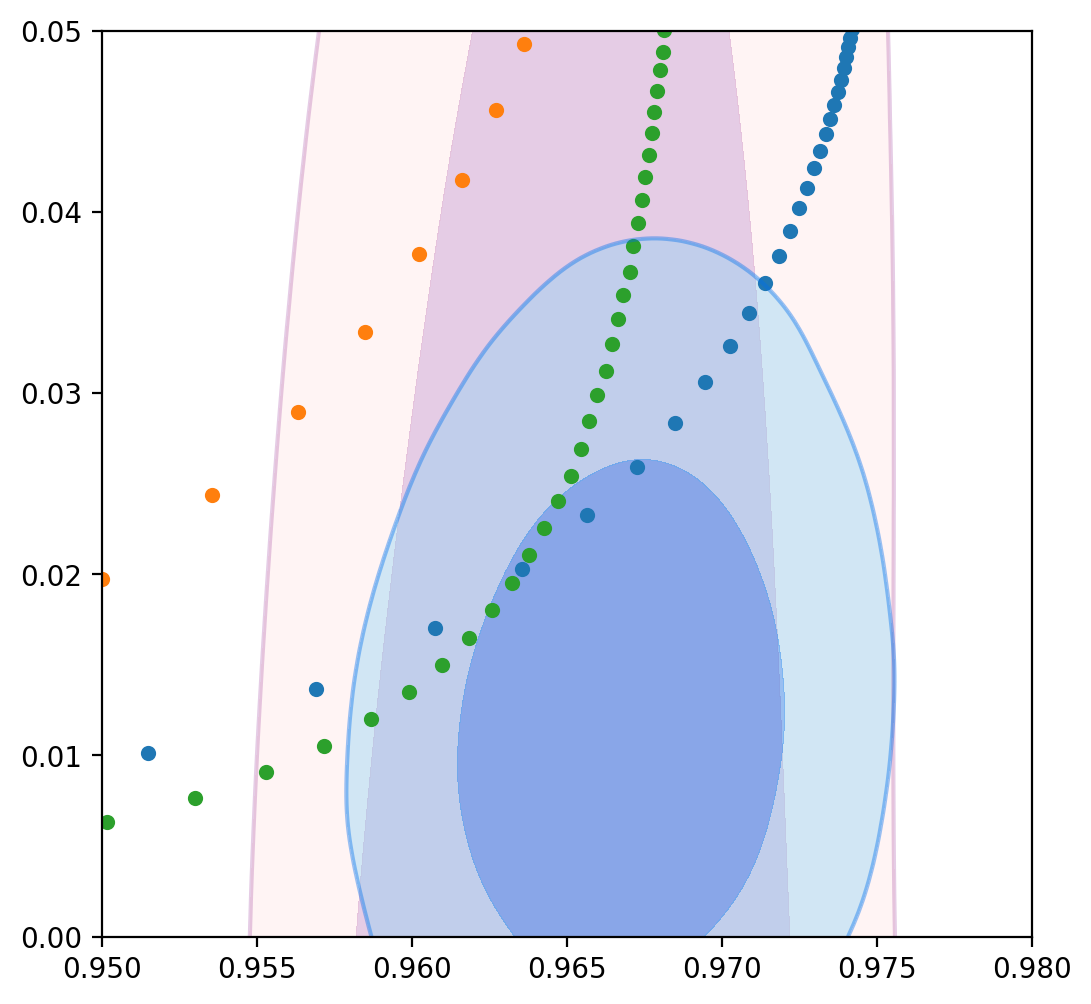}};
        \node[below] at (0.25, -4.0) {$n_s$};
        \node[rotate=90, above] at (-4.0, .15) {$r$};
    \end{tikzpicture}
    \caption{Observable predictions in the $r-n_s$ plane for the potentials depicted in Fig.~(\ref{corrected}) at $N=60$ efolds of inflation. The contours are from Planck 2018 \cite{Planck:2018vyg} and the improved Bicep/Keck 2018 constraints \cite{BICEP:2021xfz}. The blue dotted line corresponds to the uncorrected quadratic hilltop model (c.f.~Fig.~1 in \cite{Stein:2022cpk} or Fig.~(\ref{planck1}) here for the same predictions plotted on a log scale), while the yellow line corresponds to the potential with a single polynomial correction and the green lines corresponds to the potential with a series of polynomial corrections. We can see that with the potential that received a single polynomial correction the predictions for $r$ are raised, while for the potential that received a series of polynomial corrections the predictions for $r$ are lowered in part of the plane before turning sharply upwards. There is a tremendous amount of freedom in higher order terms to affect inflationary observables.}
    \label{curves_in_plane}
\end{figure}

How then do we lower $r$? Taking our cue from the suggestive results in \cite{Stein:2022cpk}, we will need to construct higher order terms that induce an earlier end to inflation. Let's briefly consider an ad-hoc example to explicitly see how one of such potential could produce a lower $r$.
\begin{equation}\label{series correction potential}
V(\phi)=V_0\left[1-\beta_2\left(\frac{\phi}{\mu}\right)^2-\beta_4\left(\frac{\phi}{\mu}\right)^4 + \beta_6\left(\frac{\phi}{\mu}\right)^6\right], 
\end{equation}
with $\phi_{vev} = \mu/\sqrt{3}$ and the coefficients $\beta_2 = 3$, $\beta_4 = 9$, and $\beta_6 = 27$ chosen to ensure $V(\phi_{vev}) = 0$. As one can see in Fig.~(\ref{corrected}), the introduction of an additional negative term causes the potential to dip more steeply and end inflation before either the quadratic hilltop model of Eq.~(\ref{hilltop potential}) or the potential with the polynomial correction at $p=4$ in Eq.~(\ref{polynomial}). However, the final correction term is positive and successfully provides a smooth end to inflation. See Fig.~(\ref{curves_in_plane}) for the $r-n_s$ curves resulting from these different hilltop potentials.

\section{Results}\label{sec:predictions}

While the above are options are straightforward ways one could think of to stabilise the potential, they are all somewhat ad-hoc. As any analytic potential admits of a Taylor expansion,
\begin{equation}
V=V_0+\left.\frac{d V}{d \phi}\right|_{\phi=0} \phi+\left.\frac{1}{2} \frac{d^2 V}{d \phi^2}\right|_{\phi=0} \phi^2+\left.\frac{1}{6} \frac{d^3 V}{d \phi^3}\right|_{\phi=0} \phi^3+\cdots,
\end{equation}
there are a huge number of possibilities that could show up at higher orders. Thinking of this as an effective field theory \cite{Burgess:2020tbq}, the relevant symmetries and energy scale cut-off will determine which of the terms show up and at what order the expansion terminates. There are many potentials whose leading order terms are quadratic and thus would be well represented at leading order by the quadratic hilltop model, but with a number of higher order terms that could perhaps serve our purposes to both stabilise the potential, as well as modify observable predictions for $r$ or $n_s$.

Thus, given our ignorance regarding inflationary energy scales and a lack of any especially compelling theoretical motivations to consider stabilising correction terms of any particular type, we will work very generally and consider models given by a polynomial expansion
\begin{equation}\label{expansion}
V(\phi) = V_0 + \sum_{n=2}^{n=q} \frac{a_n}{n!} \left(\frac{\phi}{\mu}\right)^n,
\end{equation}
where the various $a_n$'s represent expansion coefficients  of approximately $\mathcal{O}(1)$, and from now on we choose $\mu = M_{Pl}$ so that $\phi$ is scaled as $\phi \rightarrow \phi/M_{Pl}$. As depicted in Eq.~(\ref{setup}), we restrict $a_2 < 0$ in order to ensure that our leading order term is described by the quadratic hilltop model, while $a_{n}$, where $2<n<q$, can take positive or negative values, and $a_{q}$ is the order at which the expansion terminates and must be positive to ensure that the potential stabilises.
\begin{equation}\label{setup}
\begin{array}{|c|c|}
\text { coefficient } & \text { range } \\
\hline \hline a_2 & \in [-1.0,-.01] \\
a_{2<n<q} & \in [-1.0, 1.0] \\
a_q & \in [.01, 1.0]
\end{array}
\end{equation}
So, for example, going out to $q=6$ would be given by the following:
\begin{equation}\label{polynomial_expansion}
V(\phi)=V_0+\frac{a_2}{2!}\phi^2+\frac{a_3}{3!}\phi^3+\frac{a_4}{4!}\phi^4+\frac{a_5}{5!}\phi^5+\frac{a_6}{6!}\phi^6,
\end{equation}
where $a_6$ is required to be positive, $a_2$ is required to be negative, and all the $a_n$'s in between are bounded as indicated above in Eq.~(\ref{setup}). Of course, there is a tremendous amount of freedom in an expansion of this type, and the resulting potentials can take on a number of shapes, possibly having multiple minima. Thus, when we generate a potential, we then span a wide range of $\phi$ to find the global minimum and adjust $V_0$ such that $V(\phi_{vev}) = 0$ at this point (see \cite{Abel:2022nje} for a similar set-up with more general polynomials that include linear terms). Here we choose to span from $\phi \in [0, 20 M_{Pl}]$ in search of stabilising minima.

Given the high dimensionality of the parameter space and potential degeneracies between the parameters, we will pursue two strategies to explore it and assess the observational predictions of a single-field model given by an expansion of the form Eqs.~(\ref{expansion}-\ref{setup}). In Sect.~\ref{coverplane}, we will pursue a Markov chain Monte Carlo (MCMC) strategy to sample values from the parameter space to ascertain some generic features of this model. In Sect.~\ref{minimize}, we will pursue minimization strategies to see how far we truly can push $r$ with this general set up.

\subsection{Covering the $r-n_s$ plane}\label{coverplane}

We begin by exploring MCMC simulations to sample the parameter space described in Eq.~(\ref{setup}) for the potential given Eq.~(\ref{expansion}). There are a number of ways to go about this, but here we found that sampling the parameters from Eq.~(\ref{setup}) in a latin hypercube while uniformly sampling efold values in the range $N \in [50, 70]$ was effective for probing the observable parameter space. Furthermore, we restricted ourselves to viable models that have a scalar spectral index within the Planck allowed regions, which is approximately $.955 \geq n_s \leq .975$. While we could extend out to any order in the expansion given by Eq.~(\ref{expansion}), we found that that the $r-n_s$ plane rapidly becomes saturated well beyond sensitivity forecasts for future CMB experiments by the time we get to $q=6$. So, in this section, we focus on orders $q=4,5,6$.

Beginning with $q=4$, we sampled until we had obtained $\sim 10,000$ viable models; finding that at this order the observable predictions for this model have a relatively tight structure that, for the most part, predict \textit{higher} $r$ values than the uncorrected quadratic hilltop model. In light of the discussion of Sec.~(\ref{sec:hilltop}) and the results of \cite{Hoffmann:2022kod}, this is not terribly surprising. The stabilising correction term occurs at $a_4$ (a relatively high order), meaning that it will have a more significant impact on the on the potential trajectory during inflation than it would if it were a higher order term. Meanwhile, there is not that much freedom in $a_2$ or $a_3$ to induce an earlier end to inflation.

Moving to orders $q=5$ and $q=6$, we sampled in a latin hypercube until we obtained $\sim 50,000$ viable models for both $q=5$ and $q=6$ because the parameter space is significantly larger and more varied than in the $q=4$ case. We find that the predictions begin to rapidly saturate the $r-n_s$ plane, and that the $r$ values can be pushed significantly lower.\footnote{See \cite{Wolf:2023uno} for similar results and discussions in the context of quintessence driven dark energy, where the quadratic hilltop model can be shown to sweep the $w_0-w_a$ plane that parameterises dark energy phenomenology.} Again, considering the results of \cite{Hoffmann:2022kod}, this more varied behavior makes sense. The fact that the stabilising correction term is higher order means that it has less of an impact on the actual period of inflation, only coming into play at the very end where it sharply corrects and stabilises the potential. At the same time, there is more freedom in the other orders to carve out an inflationary history that lowers $r$.

\begin{figure}
    \centering
    \begin{tikzpicture}
        \node at (0, 0) {\includegraphics[width=0.45\textwidth]{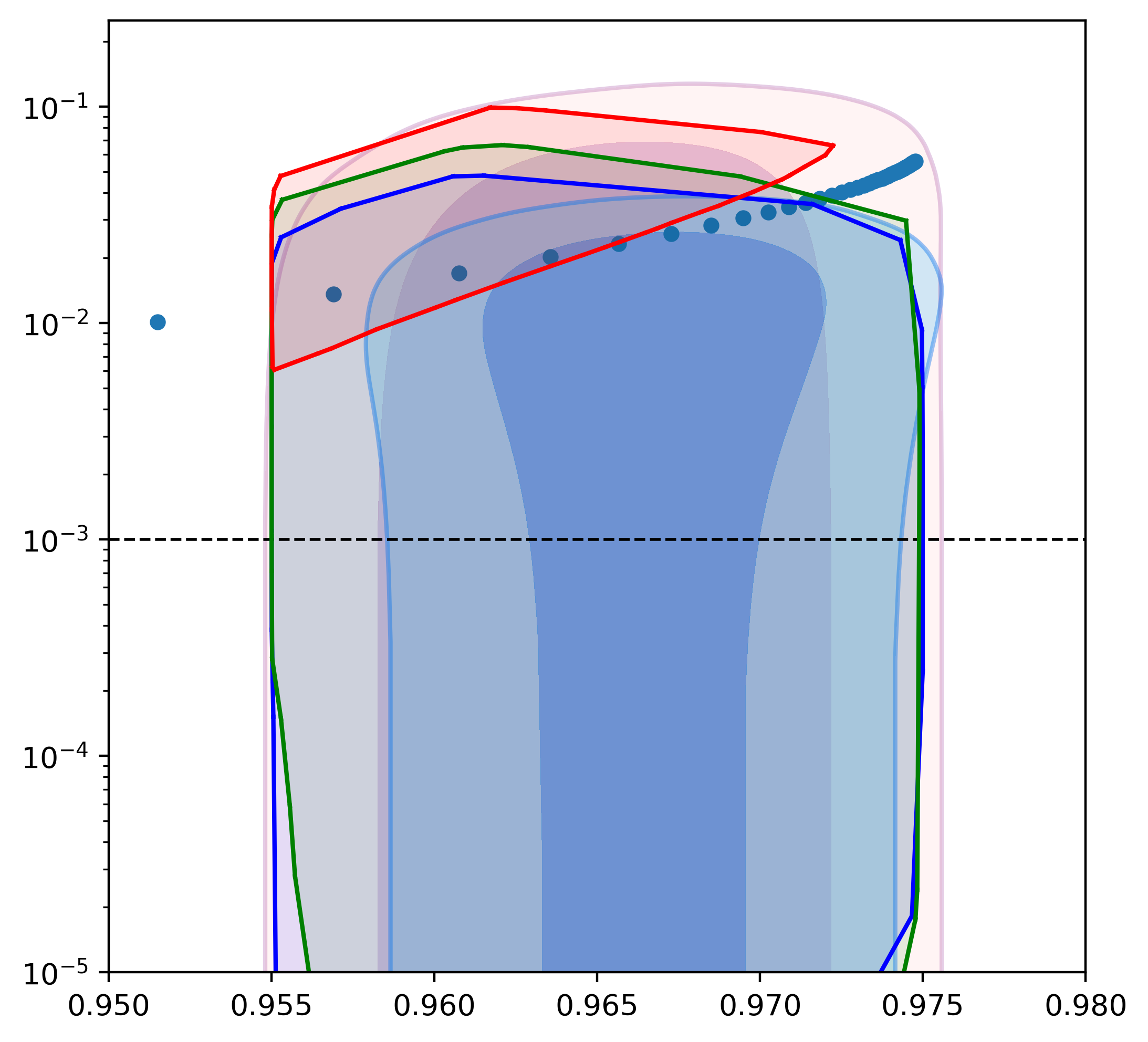}};
        \node[below] at (0.25, -4.0) {$n_s$};
        \node[rotate=90, above] at (-4.0, .15) {$r$};
    \end{tikzpicture}
    \caption{Results of MCMC sampling the inflation model described by the expansion and parameters in Eqs.~(\ref{expansion}-\ref{setup}). The red outline contains the results for $q=4$, the green outline contains the results for $q=5$, the blue outline contains the results for $q=6$, the blue dotted line represents the predictions for the uncorrected quadratic hilltop model at $N=60$ efolds of inflation (c.f.~Fig.~1 in \cite{Stein:2022cpk} or Fig.~(\ref{curves_in_plane}) here), the black dashed line represents the observational forecasts for the next generation CMB-S4 experiment which project a sensitivity of $r\sim10^{-3}$ \cite{CMB-S4:2016ple}, and the contours representing the allowed regions of parameter space are from Planck and Bicep/Keck respectively \cite{Planck:2018vyg, BICEP:2021xfz}. While the sampled space for the $q=4$ model is relatively tight, we see in going to orders $q=5$ and $q=6$ that these models rapidly begin to saturate the $r-n_s$ plane all the way down to $r\sim10^{-5}$ (and below)---roughly two orders of magnitude below the projected sensitivity of CMB-S4.}
    \label{planck1}
\end{figure}

\begin{figure}
    \centering
    \begin{tikzpicture}
        \node at (0, 0) {\includegraphics[width=0.45\textwidth]{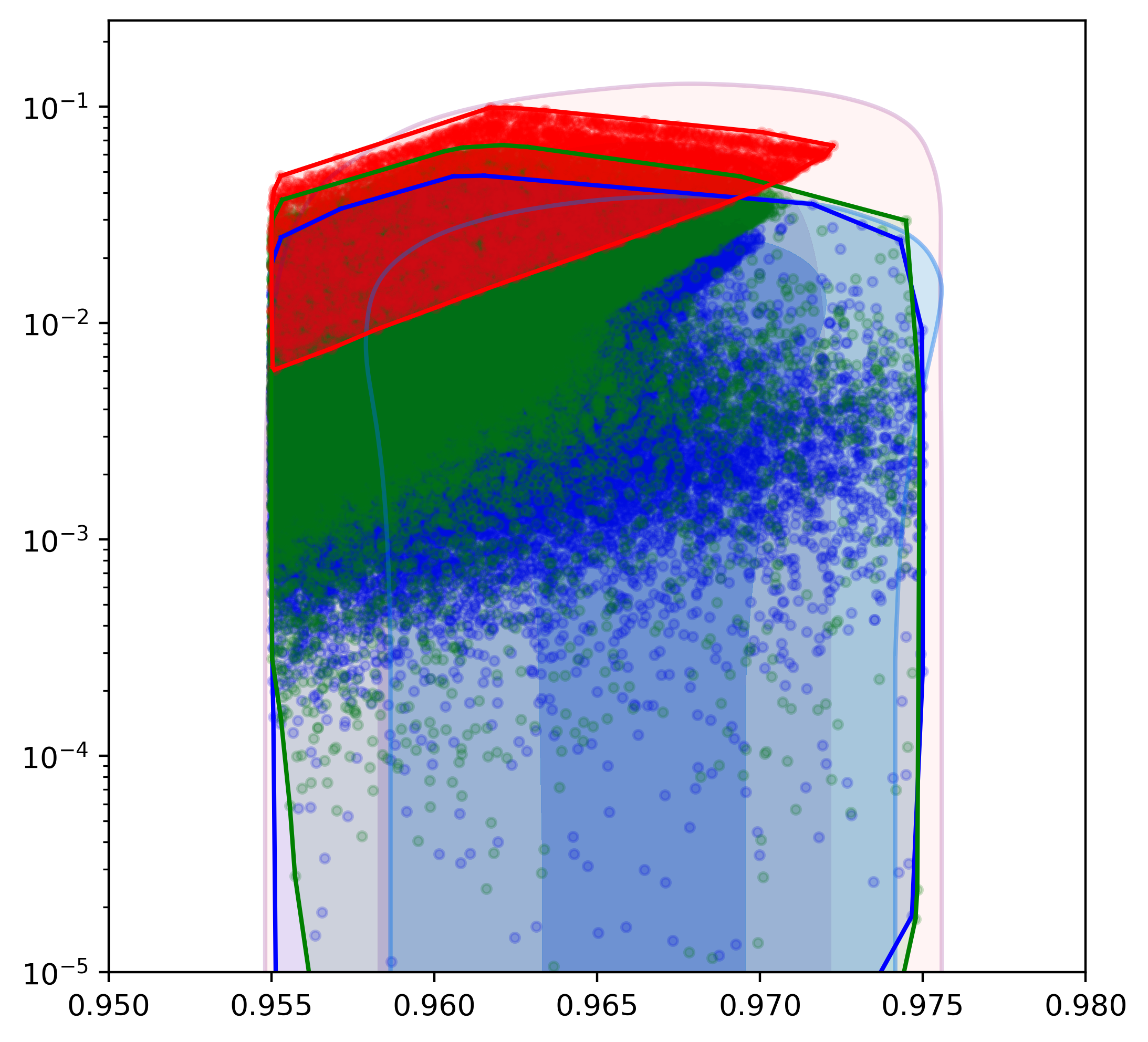}};
        \node[below] at (0.25, -4.0) {$n_s$};
        \node[rotate=90, above] at (-4.0, .15) {$r$};
    \end{tikzpicture}
    \caption{Scatter plot results of MCMC sampling the inflation model described by the expansion and parameters in Eqs.~(\ref{expansion}-\ref{setup}). As before the red represents $q=4$, the green represents $q=5$, and the blue represents $q=6$. While Fig.~(\ref{planck1}) only depicted the regions that encompassed the parameter space for the various $q$ orders in the expansion that were explored, this figure gives some insight into how the points are distributed. We see that while both $q=5$ and $q=6$ span similar areas in $n_s$ and seem to reach arbitrarily low values of $r$, $q=6$ distributes the points more evenly across the $r-n_s$ plane and contains more very small values of $r$. This is not surprising considering that having the stabilising term at a higher order will make for a sharper correction at the end that has less of an effect on the dynamics throughout the range of the potential, while also allowing for more freedom in the other parameters to lower $r$.}
    \label{planck2}
\end{figure}

These results are also significant for another reason. Sensitivity forecasts for the next generation CMB-S4 experiment project a sensitivity of $r \sim 10^{-3}$ \cite{CMB-S4:2016ple}. Here, the observable predictions for the model given by the expansion we are considering covers essentially the entire viable $n_s$ range and many orders of magnitude below the sensitivity forecasts in $r$. This is yet another suggestion, in alignment with both \cite{Stein:2022cpk} and \cite{Kallosh:2019jnl}, that ``simple'' single-field inflationary models can never be ruled out with a non-detection of $r$.\footnote{See \cite{Kallosh:2019jnl, Kallosh:2018zsi} for discussions of single-field $\alpha$-attractor and modified D-brane models which the authors show can also cover huge swaths of the $r-n_s$ plane.} These results are explored and summarized in Figs.~(\ref{planck1}-\ref{planck2}). 

\subsection{Optimizing a minimum value for $r$}\label{minimize}

While it is clear that models described by the expansion in Eq.~(\ref{expansion}) can cover the observable parameter space in the $r-n_s$ plane arbitrarily as far as foreseeable experimental sensitivity is concerned, there is still the question of how low $r$ can possibly go. In order to explore an answer to this question, we will go order by order Eq.~(\ref{expansion}), and optimize the potential $V(\phi)$ given the parameter space described by Eq.~(\ref{setup}) in order to minimize the value of $r$.

There are a variety of tools one could utilize here. One of the more obvious strategies to pursue is to deploy something like SciPy's minimize function. This tool takes an ``objective function'' that can receive a number of variables as an input and outputs a scalar value, and then finds a minimum value for the objective function. In our case, the objective function takes the potential in Eq.~(\ref{expansion}) and parameter bounds in Eq.~(\ref{setup}) as inputs, and then outputs the calculated values of $r$ and $n_s$. While this minimization tool has a number of methods available, including both gradient-based and derivative-free methods, we found that utilising this approach tended to frequently get stuck in local minima that would not always produce $r$ values as low as some of the ones we would find in the MCMC simulations. We attribute this to the inherent structural complexity of the function and the vast size of the parameter space. Gradient-based methods would quite naturally get stuck because they are based on first derivatives and this function clearly has a vast landscape of local minima. However, even the derivative-free methods would also seem to find themselves eventually getting stuck in local minima. 

After exploring this option, we then deployed more sophisticated optimization techniques, settling on SciPy's differential evolution algorithm, which is specifically designed to find the global minima of non-convex functions with possibly many local minima. Like minimize, it optimizes some object function with input variables and a scalar output; however, it is also a population-based approach that maintains a number of candidate solutions which iteratively evolve through population crossover, mutation, and selection. This allows the optimization to more effectively explore the parameter space even if the objective function contains many local minima. Here, we found that population sizes of around $25-50$ (initialized from a latin hypercube) along with relatively high mutation ($\sim 1-2$) and recombination/crossover rates ($\sim 1$) was effective for exploring the parameter space. 

\begin{figure}
    \centering
    \begin{tikzpicture}
        \node at (0, 0) {\includegraphics[width=0.45\textwidth]{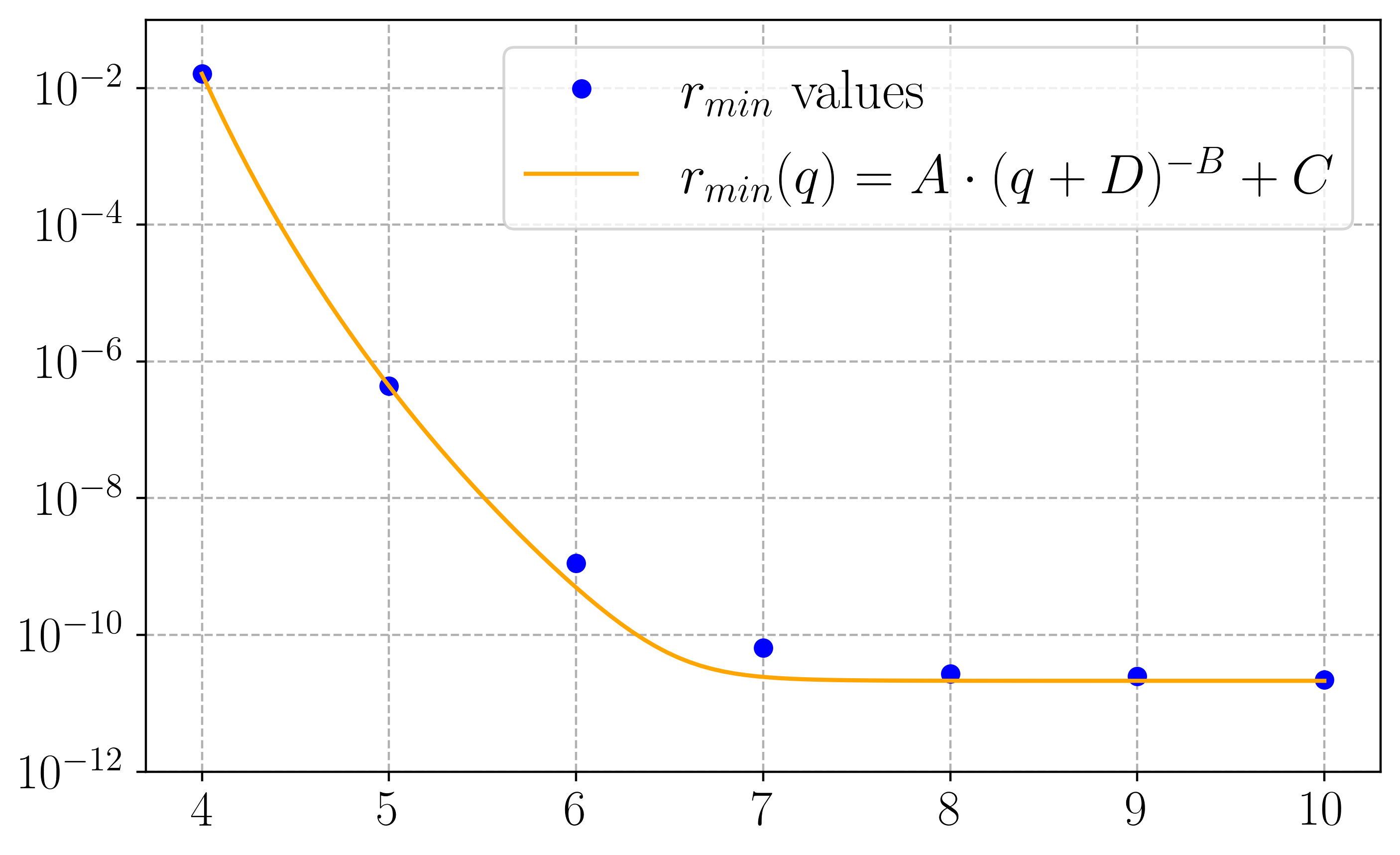}};
        \node[below] at (0.25, -2.5) {$q$};
        \node[rotate=90, above] at (-4.0, .15) {$r$};
    \end{tikzpicture}
    \caption{Results of the optimization procedure to minimize $r$ order by order in $q$. As one can see by inspection or by Eq.~(\ref{eq:curvefit}), the relationship is well-described by a power law that asymptotes at $r\sim 10^{-11}$. See Table.~(\ref{tab:params}) for exact $r_{min}$ values.}
    \label{rmin}
\end{figure}

In the following results, for reasons that will soon become apparent, we went even further out in terms of the order of the potential's expansion than we did in the MCMC sampling. We found that the higher order one caries out the expansion in Eq.~(\ref{expansion}) to, the lower the mininum $r$, but that by $q=8$ the r values seem to asymptote. We then carried out the expansion all the way out to order $q=10$ to confirm that this is the case. As one can see from examining Fig.~(\ref{rmin}) or Table.~(\ref{tab:params}), the optimization procedure we employed produces a smooth curve that shows the minimum tensor/scalar ration $r_{min}$ obtained from the potential $V(\phi)$ as a function of the order $q$ at which the expansion in Eq.~(\ref{expansion}) was carried out to. The functional relationship is well-described by a power law $r_{min} \propto q^{-B}$, before it asymptotes at a lower bound of $r_{min}\sim10^{-11}$. The exact functional form is given by:
\begin{equation}\label{eq:curvefit}
r_{min} (q) = A \cdot (q + D)^{-B} + C,
\end{equation}
where $A \approx 17.77$, $B \approx 19.78$, $C \approx 2.13 \times 10^{-11}$, $D \approx -2.58$. Taken together, it does indeed seem that for all intents and purposes, higher order terms in the expansion given by Eq.~(\ref{expansion}) effectively have the freedom to lower $r$ arbitrarily as suggested by \cite{Stein:2022cpk}.

\begin{table}[h!]
    \centering
    \begin{tabular}{|c|c|c|}
        \hline
        \textbf{\( q \)} & \textbf{\( r \)} & \textbf{\( n_s \)} \\ \hline
        4 & $1.62 \times 10^{-2}$ & 0.9550 \\ \hline
        5 & $4.378 \times 10^{-7}$ & 0.9552 \\ \hline
        6 & $1.116 \times 10^{-9}$ & 0.9606 \\ \hline
        7 & $6.456 \times 10^{-11}$ & 0.9675 \\ \hline
        8 & $2.718 \times 10^{-11}$ & 0.9551 \\ \hline
        9 & $2.469 \times 10^{-11}$ & 0.9555 \\ \hline
        10 & $2.222 \times 10^{-11}$ & 0.9570 \\ \hline
    \end{tabular}
    \caption{Table of $q$, $r$, and $n_s$ values corresponding the to results depicted in Fig.~(\ref{rmin}).}
    \label{tab:params}
\end{table}

\section{Conclusion}\label{sec:conclusion}

Throughout the literature on inflation there has been much discussion concerning what the inflationary paradigm generically predicts for crucial observables such as $r$ and $n_s$. While many of the simplest and most commonly explored models do predict a relatively high $r$ (and some have been eliminated as a result), there are now a number of counter examples that show that single-field inflation can also produce an undetectably negligible tensor/scalar ratio \cite{Stein:2022cpk, Kallosh:2018zsi, Kallosh:2019jnl}. Even going beyond our most optimistic scenarios for observational sensitivities though, studies such as this one which suggests that single-field inflation can produce tensor/scalar ratios as low as $r \sim 10^{-11}$, \cite{Lorenzoni:2024krn} which investigates a two-field model of natural inflation that can produce tensor/scalar ratios as low as $r \sim 10^{-15}$, or \cite{Braglia:2020fms} which studies the mechanisms by which additional fields can suppress $r$, underscore the almost infinite flexibility of the inflation paradigm.

Here, we explored modifications to the quadratic hilltop model and explicitly showed that introducing possible correction terms can radically alter the observable predictions, where the form of the correction terms we considered was motivated by a widely applicable effective field theory operator expansion. Such higher order correction terms are both necessary to stabilise the potential in order to ensure a smooth end to inflation, and also represent additional degrees of freedom that can lower $r$ significantly. We showed this to be the case both by randomly sampling many possible viable models described by such an expansion, as well as by optimizing to find the minimum allowed values of $r$.

Where does this leave us? A null detection of $r$ seemingly will never totally rule out single-field inflation. Yet, as these examples suggest, the further down $r$ goes the more structurally complicated the models need to become \cite{Sousa:2023unz, Boyle:2005ug}. While such models may indeed be simple in the sense that they can be described by a single canonical scalar field minimally coupled to gravity, pushing $r$ lower does indeed seem to require more structural complexity. Like dark energy \cite{Wolf:2023uno}, early universe models suffer from surprisingly stubborn underdetermination problems; however, future observations will no doubt continue to offer valuable information concerning the physical processes underlying cosmological phenomena.

\section*{Acknowledgements}

I am very grateful to Pedro Ferreira and David Sloan for many illuminating discussions. I am also grateful for financial support from St.~Cross College, Oxford.



\newpage
\bibliography{refs}

\end{document}